\documentclass[10pt,conference]{IEEEtran}

\IEEEoverridecommandlockouts
    \usepackage[pdftex]{graphicx}
    \graphicspath{{../pdf/}{../jpeg/}}
    \DeclareGraphicsExtensions{.pdf,.jpeg,.png}  
    \usepackage[cmex10]{amsmath}
    \usepackage{hyperref}
    \usepackage{algpseudocode}
    \usepackage{array}
    \usepackage{mdwtab}
    \usepackage{eqparbox}
    \usepackage{url}
    \usepackage{comment}
    \usepackage{subfigure}
    \usepackage{romannum}
    \usepackage{amssymb}
   \usepackage{algorithm} 

    \usepackage[utf8]{inputenc}
    \usepackage{amsmath,amssymb}
    \usepackage{algorithm}
    \usepackage{algpseudocode}

    \usepackage{graphicx} 
    \usepackage{booktabs} 
    \usepackage{cite}
\usepackage{caption, multirow, makecell}
\usepackage{siunitx}
\usepackage{mhchem}
\usepackage{float}
\usepackage{adjustbox}
    \usepackage{multirow}

\usepackage[T1]{fontenc}

\usepackage{tabularx} 
\IEEEoverridecommandlockouts

\begin{document}

\title{\textit{ProtoBeam}:  Generalizing Deep Beam Prediction to Unseen Antennas using Prototypical Networks}

\author{\IEEEauthorblockN{ Omar Mashaal\IEEEauthorrefmark{3},
Elsayed Mohammed\IEEEauthorrefmark{3},
Alec Digby\IEEEauthorrefmark{1},
Lorne Swersky\IEEEauthorrefmark{1},
Ashkan Eshaghbeigi\IEEEauthorrefmark{1},
Hatem Abou-Zeid\IEEEauthorrefmark{3}
}
\IEEEauthorblockA{\IEEEauthorrefmark{3}{Department of Electrical and Software Engineering}, 
{University of Calgary}, Canada}
\IEEEauthorblockA{\IEEEauthorrefmark{1}{Qoherent Inc.}, 
{Toronto, Ontario, Canada}} 
\thanks{The authors would like to thank Qoherent Inc. and  MITACS Accelerate for their support of this research. The authors would also like to thank Denvr Dataworks, Calgary, Canada for their high-performance compute.}
}

\maketitle

\begin{abstract}

Deep learning (DL) techniques have recently emerged to efficiently  manage mmWave beam transmissions without requiring time consuming beam sweeping strategies. A fundamental challenge in these methods is their dependency on hardware-specific training data and their limited ability to generalize. Large drops in performance are reported in literature when DL models trained in one antenna environment are applied in another.  This paper proposes the application of Prototypical Networks to address this challenge \textendash{}  and utilizes the DeepBeam real-world dataset \cite{DB} to validate the developed solutions. Prototypical Networks (PN) excel in extracting features to establish class-specific prototypes during the training, resulting in precise embeddings that encapsulate the defining features of the data.
We demonstrate the effectiveness of PN to enable  generalization of deep beam predictors across
unseen antennas. Our approach, which integrates data normalization and prototype normalization with the PN, achieves an average beam classification accuracy of 74.11\% when trained and tested on different antenna  datasets. This is an improvement of 398\% compared to baseline performances reported in literature that do not account for such domain shifts. 
To the best of our knowledge, this work represents the first demonstration of the value of Prototypical Networks for domain adaptation in wireless networks, providing a foundation for future research in this area.

\end{abstract}

\IEEEoverridecommandlockouts

\IEEEpeerreviewmaketitle


\section{Introduction}

In the realm of wireless communication, machine learning models are traditionally developed with the expectation that the training and testing distributions are identical.
However, this simplification often proves inadequate in practical scenarios, leading to significant performance degradation due to distribution shifts in new, unseen environments. These shifts primarily arise because of inherent hardware differences, antenna properties, and propagation environments utilized during training fail to represent those encountered in actual deployment. This discrepancy is particularly pronounced in mmWave beam management applications, where precise beam alignment is crucial for system performance, highlighting a gap between theoretical assumptions and operational realities.

To fully exploit the bandwidth potential of mmWave technologies for next-generation wireless networks, efficient beam management is essential for robust connectivity and signal transmission. Deep learning (DL) has emerged as a powerful tool for refining beam management strategies, enabling dynamic beam prediction and alignment  \cite{DB, L1,MM2}. DeepBeam \cite{DB} leverages deep learning to optimize beam selection using I/Q data. This framework is designed to infer the Angle of Arrival (AoA) and identify the beam used by the transmitter through passive reception of ongoing transmissions. It offers operational flexibility, functioning with both single and multiple RF chains. This capability allows DeepBeam to bypass traditional beam sweeping, thereby streamlining the beam management process. Despite these capabilities, DeepBeam encounters challenges with classification accuracy when models trained in one antenna environment are applied in another. This underscores the importance of domain adaptation techniques to preserve the efficacy of DL models amidst mmWave system and hardware variability.

\begin{figure}[t!] 
\centering
\includegraphics[width=3.6in, height=2in, keepaspectratio]{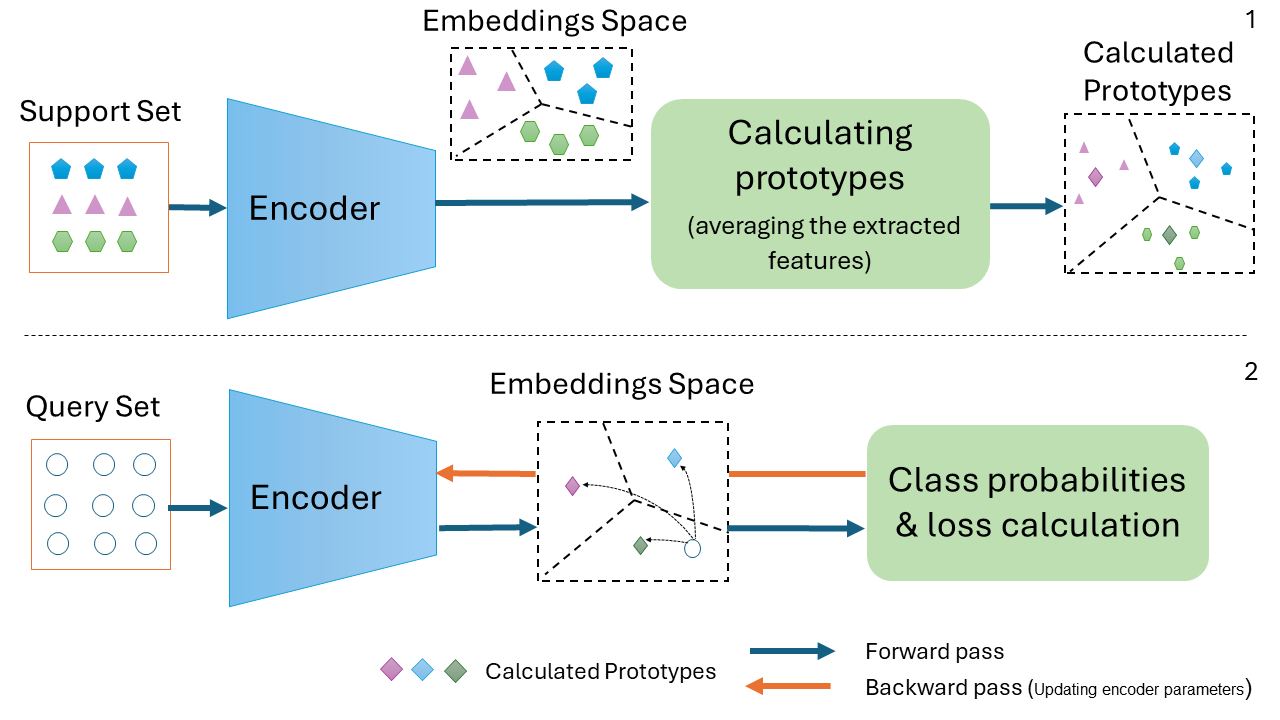}
\caption{Prototypical Network Architecture and Training \cite{DA2}.}
\label{protoFig}
\end{figure}

Several advances toward such \emph{domain generalization} are actively being researched by the DL community. 
This includes utilizing strategies such as Prototypical Networks (PN) \cite{DA2} and meta-learning techniques like Model-Agnostic Meta-Learning (MAML) \cite{DA1}. Prototypical Networks are distinguished by their capacity to adapt to new domains without additional training. Unlike MAML, which necessitates model fine-tuning, PNs excel in extracting features to establish class-specific prototypes during the initial training phase. These prototypes, the class averages of embedded support examples as depicted in Figure \ref{protoFig}, facilitate new example classification by measuring distances to the prototypes. This method of feature extraction and prototype creation allows PNs to navigate domain shifts effectively. We hypothesize that this makes them well-suited for tasks like beam classification in mmWave communication \cite{DB}, where generalizing from limited data is vital.

In response to the challenges of domain shifts in mmWave beam prediction, our paper aims to propose and investigate the application of Prototypical Networks to address these limitations. 
The primary contributions of our paper are:
\begin{itemize}
\item We propose and develop ProtoBeam \textendash{} an approach using Prototypical Networks (PN) for mmWave beam classification. ProtoBeam enables DL models to be deployed on unseen antenna setups without requiring retraining. To the best of our knowledge, this work represents the first demonstration of the value of PN for domain adaptation in wireless networks, providing a foundation for future research in this area.  

\item We develop and integrate data augmentation, data normalization, and prototype normalization methods to improve the performance of ProtoBeam. We show that these techniques improve accuracy by over 20\% and are essential for effective PN design in beam classification. Our analysis includes examining the effects of various design choices, specifically focusing on the number of training shots used in beam classification.

\item We utilizes the DeepBeam real-world dataset \cite{DB} that is characterized by diverse antenna configurations to demonstrate the effectiveness of ProtoBeam. 
Our approach achieves an average beam classification accuracy of 74.11\% when trained and tested on different antenna datasets. This is an improvement of 398\% compared to baseline performances reported in literature.

\end{itemize}

The results of this paper highlight the significant potential that Prototypical Networks have to address the domain shift challenge encountered in real-world AI-powered wireless communications. All the code to reproduce our results and conduct further research in this area is available at \url{https://github.com/omarmshaal/ProtoBeam} The rest of the paper is organized as follows: Section \Romannum{2} summarises related work. Section \Romannum{3} describes our methodology and the dataset used. Section \Romannum{4} discusses our results and findings. We conclude the paper in Section \Romannum{5}.


\section{Related Work}

Meta-learning, as a solution to domain shift challenges, is increasingly attracting attention in the wireless domain, with several notable works exploring its application in beam management. For instance, the work in \cite{RW1} introduces a meta-learning algorithm designed for beam prediction across sub-6GHz and mmWave bands, leveraging sub-6GHz CSI to navigate distribution shifts through a bi-level optimization meta-learner. Additionally, \cite{RW2} employs MAML and transfer learning to adapt to domain shifts in beamforming, demonstrating that MAML outperforms transfer learning for predicting the power allocation vector. Furthermore, \cite{RW3} employs a few-shot MAML approach for classifying incident wave angles using received signal strength (RSS) in air-to-air networks, adapting to dynamic 3D channel conditions with limited data availability. The work in \cite{RW4} aims to streamline the beam selection process by substituting exhaustive searches with probing and employs the MAML framework to swiftly adapt to new settings. Lastly, \cite{RW5} proposes an adaptive beamforming method that utilizes support vector regression (SVR) for fast adaptation. This method involves using a pre-trained deep neural network as a feature extractor, with only the SVR model requiring training during the adaptation stage to quickly adjust to new environments.                                                                                               
While previous approaches in beam management domain adaptation primarily employ MAML for tasks such as power vector allocation and beam classification, these methods necessitate model retraining or fine-tuning, adding extra computational complexity and time. In contrast, our study utilizes Prototypical Networks for mmWave beam classification, aiming to handle domain shifts across antennas without requiring retraining. This approach seeks to avoid extensive training phases common in methods like MAML, targeting classification directly. Our beam management application focus is also different from prior work. We aim to identify the beam used by the transmitter through \emph{passive} reception of ongoing transmissions, building on the real-world data measurements for different antennas presented in \cite{DB}.


\section{ProtoBeam: Generalizable Deep Beam Prediction for mmWave}

In this section, we first present the domain adaptation problem encountered when DL beam classification models are trained and then deployed on different antenna hardware. We then present our ProtoBeam framework, showcasing how Prototypical Networks are employed to handle such domain shifts in mmWave beam management systems.

\subsection{Problem Formulation}

\begin{figure}[!tbh]
  \centering
  \subfigure[$ $]{\includegraphics[trim={1cm 1cm 1cm 1cm}, width=0.43\textwidth]{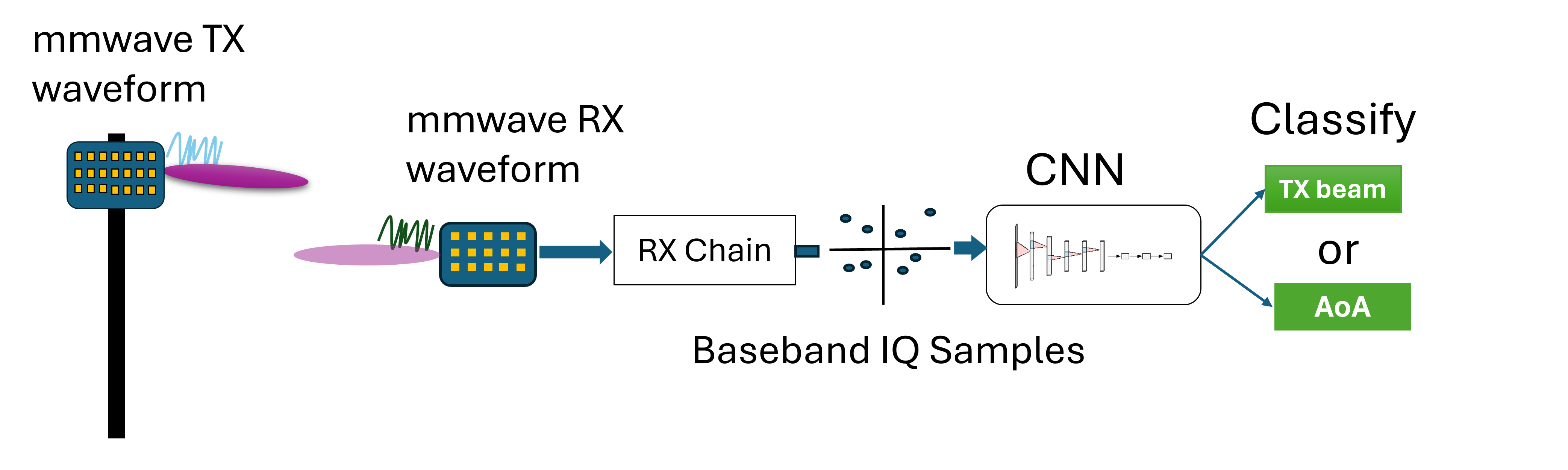} \label{deepbeam}}
   \subfigure[$ $]{\includegraphics[width=0.43\textwidth]{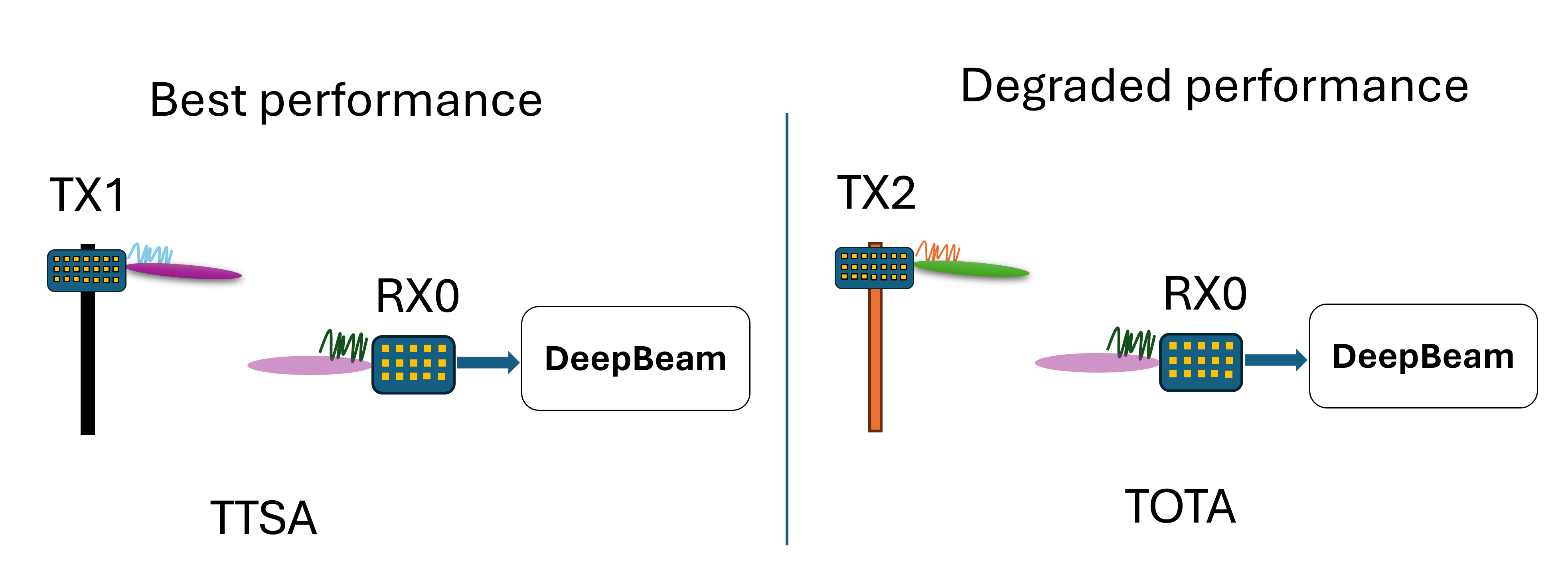} \label{domainadapt}}
  \caption{(a) Overview of the DeepBeam framework and (b) The TOTA domain adaptation challenge in beam prediction.}
  \label{architecture_diagram}
\end{figure}

DeepBeam \cite{DB} is the first framework designed for mmWave networks that utilizes waveform-level deep learning to manage beams without requiring explicit coordination between transmitters (TX) and receivers (RX). This approach allows DeepBeam to leverage ongoing transmissions to infer the AoA and determine which transmit beam (TXB), from a certain codebook, is being used by the TX. Once enough transmissions are eavesdropped, DeepBeam can recommend the optimal beam for communication by ranking the inferred TXBs by their associated Reference Signal Received Power (RSRP), eliminating the need for pilot signals. Additionally, DeepBeam is directly connected to the device's RF chain as shown in Figure \ref{deepbeam}, allowing it to process I/Q samples at the physical (PHY) layer without requiring synchronization between the transmitter and receiver.

Despite the capabilities of DeepBeam, the framework encounters challenges with domain shifts, particularly evident when deploying the 24-beam codebook. For instance, the average accuracy for \textit{Training and Testing On the Same Antenna} (TTSA) stands at 78.51\%. However, this accuracy significantly decreases when the system is subjected to a \textit{Training on One Antenna and Testing on Another} (TOTA) scenario, where the training data might be collected by one set of RF hardware (such as TX1 and TX0), and then tested by another (such as TX2 and TX0) as shown in Figure \ref{domainadapt}. TOTA results in an accuracy drop to 16.97\%. Furthermore, when training and testing involve a mixed dataset with waveforms from all antennas, accuracy improves to 49.41\%, yet still substantially lags behind the TTSA scenario. Therefore, the core challenge is to enhance the model's generalization capabilities to maintain high TTSA accuracy while substantially improving TOTA accuracy in diverse antenna setups. This will ensure more reliable beam classification across different mmWave hardware and system capabilities.

\subsection{Dataset}

The DeepBeam dataset\cite{DB}, utilized in our study, captures experimental data from 24 horizontal beam codebooks, organized into HDF5 files. Each file represents a combination of transmission and reception antenna variations, specifically using four different SiBeam 60 GHz frontends, numbered TX0, TX1, TX2, and TX3. These setups include I/Q samples across three receiver gain settings, covering SNR levels from -15 dB to 20 dB. Each transmission antenna and receiver gain combination comprises 150,000 blocks, with each block containing 2048 I/Q samples. In total, the single file of the dataset includes 10,800,000 blocks for the 24 beam configurations across all gain settings, offering a robust volume of data for in-depth analysis under a variety of signal conditions.

\subsection{Prototypical Networks for Cross-Antenna Beam Prediction}

Prototypical Networks operate on the principle that data points within an embedding space cluster around a central prototype for each class as shown in Figure \ref{protoFig}. Utilizing an encoder for non-linear mapping, the networks cluster features around a central prototype for each class within an embedding space. Classification becomes a nearest-neighbour problem, with the class of a query point determined by its proximity to these prototypes. 
ProtoBeam employs this concept of Prototypical Networks for beam classification where prototypes are created during training using a training set of a particular antenna.
 
The adaptive learning mechanism of Prototypical Networks is then expected to deliver stable classification performance \emph{across various antenna setups}, including those not encountered during training. Details are discussed next.

\subsection{Prototypical Network Architecture and Training}

In our training strategy for the ProtoBeam PN, we employ a k-way classification method. Here, 'k' denotes the number of classes randomly selected from the available 24 beam classes for each training episode, which establishes our sampling framework. We also implement a k-shot approach, where a specific number of examples from each chosen class are selected to compose the support set for that episode. The dual processes of random class selection and support set assembly are vital components of each training episode. These steps are depicted in Figure \ref{protoFig} and are detailed in Algorithm \ref{alg:train_protoBeam}.

\subsubsection{Prototype Calculation}

We initiate the training episode by selecting a number of beams (\( N_B \)) and a corresponding number of I/Q samples per beam from the dataset, which form the support set. Beam prototypes are computed by embedding the I/Q samples into a feature space and averaging these embeddings, as in Figure  \ref{protoFig} step 1, which is also defined by the equation:

\begin{equation}
p_b = \frac{1}{|I_b|} \sum_{(x_i, y_i) \in I_b} f_\phi (x_i)
\end{equation}

where \( I_b \) denotes the set of I/Q samples for the target beam \( b \), \( x_i \in \mathbb{R}^D \) is the feature vector of the I/Q sample with dimension \( D \) being \( (2 \times 2048) \), \( y_i \) is the corresponding label for the target beam, with the total number of beams (\( B \)) equal to 24, \( p_b \) is the prototype for the target beam \( b \), and \( f_\phi \) is the encoder function that maps the I/Q samples \( x_i \) to the embedding space.

\subsubsection{Encoder}
In our implementation, we employed DenseNet \cite{densnet} for the encoder (feature extractor) due to its effectiveness in promoting feature reuse and mitigating overfitting. The architecture comprises three dense blocks, each containing 5 layers, tailored to balance depth with computational efficiency. The output size of the encoder, which determines the size of the prototype, is set to 128. While alternative models could potentially serve this domain, our exploration is confined to DenseNet's application within the current study's scope.

\subsubsection{Classification and Loss Computation}

During the classification phase, the network calculates the probability that a query I/Q sample \( x \) belongs to a specific beam \( b \) by applying a softmax function over the Euclidean distances from the query's embedding to each beam prototype, shown in Figure \ref{protoFig} step 2. This probability is given by:

\begin{equation}
p_\phi(y = b|x) = \frac{\exp(-d(f_\phi(x), p_b))}{\sum_{b'} \exp(-d(f_\phi(x), p_{b'}))}
\end{equation}

where \( d \) is the chosen distance metric, \( f_\phi \) is the encoder function producing the embedding, \( p_b \) is the prototype for the target beam \( b \), and the sum in the denominator extends over all beam prototypes \( b' \).

The beam with the highest probability \( p_\phi(y = b|x) \) is predicted as the label for the query I/Q sample \( x \). The loss for each query sample is computed using the negative log-likelihood of the true beam label, guiding the model update during backpropagation:

\begin{equation}
\mathcal{L} = -\log p_\phi(y = y_{\text{true}}|x)
\end{equation}

This loss, averaged over all the query I/Q samples, is utilized in a gradient descent step to refine the model parameters, aiming to reduce the loss and thus boost the network's beam classification accuracy. By iteratively training and reducing loss, the Prototypical Network hones its ability to produce precise embeddings that encapsulate the defining features of each beam, improving generalization and adaptability across varying beam conditions.

\begin{algorithm}[!b]
\caption{Proposed ProtoBeam Training Algorithm}
\label{alg:train_protoBeam}

\textbf{Inputs:} {Training set $\mathcal{D} = \{(x_1, y_1), \ldots, (x_{n_b}, y_{n_b})\}$, where $x_i$ is the I/Q sample and $y_i \in \{1, \ldots, B\}$ is the target beam.\\ $\mathcal{D}_b$ denotes the subset of $\mathcal{D}$ containing all elements $(x_i, y_i)$ for target beam $b$.}\\
\textbf{Parameters:} {$n_b$ is the number of baseband I/Q samples.\\ 
$B$ is the number of target beams, $N_B \leq B$ is the number of target beams per episode. \\
$n_S$ is the number of I/Q support examples per target beam.\\ $n_Q$ is the number of I/Q query examples per target beam. \\ 
$RandSample(\mathcal{S}, N)$ denotes a set of $N$ elements chosen uniformly at random from set $\mathcal{S}$, without replacement.}\\
\textbf{Output:} {Updated model parameters after backpropagation.}

\begin{algorithmic}
\Procedure{TrainProtoBeam }{$\mathcal{D}$}
\newline \hspace*{.45cm} Select indices for target beams in this episode
\State $V \gets RandSample(\{1, \ldots, B\}, N_B)$ 
\For{$b \in V$}
    \State $I_b \gets RandSample(\mathcal{D}_b, n_S)$ \Comment{ Support}
    \State $Q_b \gets RandSample(\mathcal{D}_b \setminus I_b, n_Q)$ \Comment{Query}
    \State $p_b \gets \frac{1}{n_S} \sum_{(x_i, y_i) \in I_b} f_\phi (x_i)$ \Comment{Compute Prototypes}
\EndFor
\State $L \gets 0$ \Comment{Initialize loss for this episode} 
\For{$b \in V$}
    \For{$(x_i, y_i) \in Q_b$}
        
        \State $L \gets L + \frac{1}{N_B n_Q} \biggl[ d(f_\phi (x_i), p_b) +$
        \State $\quad \log \biggl( \sum_{b'} \exp\bigl(-d(f_\phi (x_i), p_{b'})\bigr) \biggr) \biggr]$ 
    \EndFor
\EndFor
\State Compute gradients of $L$ w.r.t. model parameters $\phi$
\State Perform backpropagation to update model parameters
\State $\phi \gets \phi - \alpha \cdot \nabla_\phi L$ \Comment{Update $\phi$ with learning rate $\alpha$}
\State \textbf{return} $\phi$ 
\EndProcedure
\end{algorithmic}
\end{algorithm}

\subsection{Data Augmentations and Prototype Normalization}

Our initial applications of Prototypical Networks to adapt to domain shifts in mmWave beam management revealed suboptimal results. To address this, we implemented data normalization and data augmentation, techniques not previously applied in DeepBeam, and also applied prototype normalization as recommended by \cite{DA1}. These enhancements were aimed at improving model robustness and accuracy. The effectiveness of these methods and their impact on performance are detailed in the Results and Discussion section.

\textbf{Data Normalization:}
We applied min-max normalization to the I/Q data to standardize input ranges, enhancing the model's robustness against variations in input scales, as follows:
\begin{equation}
x'_i = 2 \left(\frac{x_i - x_{\text{min}}}{x_{\text{max}} - x_{\text{min}}}\right) - 1
\end{equation}
where \(x_i\) is the original data value, \(x_{\text{min}}\) and \(x_{\text{max}}\) are the dataset's minimum and maximum values, respectively, and \(x'_i\) is the normalized value, scaled to the range \([-1, 1]\).

\textbf{Data Augmentation:}
We implemented random phase rotation and scaling of I/Q data as augmentation techniques to mimic real-world signal variations, enhancing the model's adaptability across diverse conditions.

\section{Results and Discussion}

This section evaluates the performance of our proposed ProtoBeam approach for mmWave beam classification. Our analysis covers several aspects: an examination of the impact of various enhancement techniques, comparative accuracy assessments across different antenna setups, and detailed evaluations of model performance through confusion matrices. While we delve into the effectiveness of individual enhancements, our primary focus remains on analyzing the model’s classification accuracy and its ability to generalize across different antenna configurations.

\textit{ProtoBeam Model Training}:
The Prototypical Network, leveraging DenseNet for feature extraction, underwent training using a 5-way, 4-shots approach until the model stopped improving. Training was executed using an AdamW optimizer, starting with a learning rate of 0.002. A learning rate schedule, scaling by a factor of 0.1, was applied as necessary to fine-tune performance. Training and test datasets were both shuffled to prevent sequence biases and corrupted samples, ensuring the integrity of prototype construction and evaluation. The PyTorch framework was used to develop and train our models.

\textit{Hardware and Software Configuration}: The methods were executed on a high-performance computing setup, comprised of a 12th Gen Intel(R) Core(TM) i9-12900K processor operating at 3.20 GHz, coupled with 64.0 GB of RAM, and powered by an NVIDIA GeForce RTX 3080 Ti graphics card. 

\subsection{ProtoBeam Accuracy}

\begin{figure}[!t]
  \centering

 {\includegraphics[trim={0.4cm 0 0.cm 0cm },clip,width=0.4\textwidth]{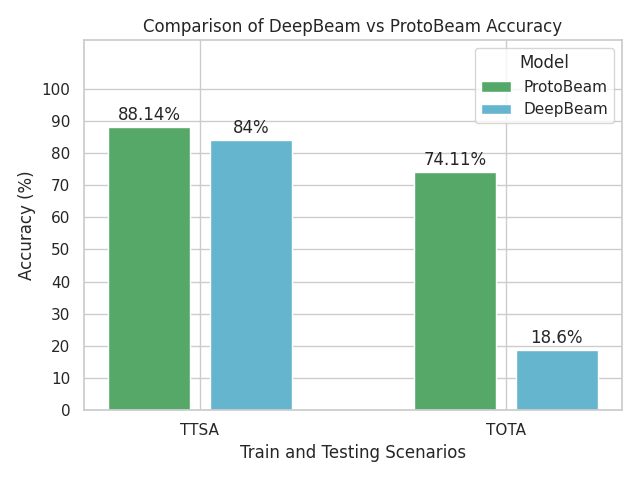} } 
  
  \caption{Accuracy Improvements of ProtoBeam.}
  \label{stackedBar1}
\end{figure}

Figure \ref{stackedBar1} displays the performance results of our Prototypical Network compared to the baseline developed in \cite{DB} that was not designed to adapt across antennas. We show separate performance bars for the case of TTSA and TOTA. In TOTA, training is on TX0 and testing is conducted on three different antennas: TX1, TX2, and TX3. The main objective of ProtoBeam is to generalize and mitigate the performance loss in the TOTA case. 
In this  scenario, ProtoBeam achieves a 74.11\% accuracy at 16-shots, a large improvement by 398.4\% compared to the baseline performance of 18.6\%. It also exceeds the mixed data training accuracy across multiple antennas by 149.9\%, which stands at 49.41\%. ProtoBeam also outperformed the baseline in the TTSA scenario, indicating the the features extracted through the Prototype Network are also effective in improving the accuracy in general. These enhancements demonstrate the ProtoBeam's robust adaptability to domain shifts, as previously highlighted by the challenges observed in the DeepBeam dataset. 
In what follows, we conduct a detailed analysis of various design effects on performance of ProtoBeam.

\begin{table}[b]
\renewcommand{\arraystretch}{1.2} 
    \centering
    \caption{ProtoBeam Performance with Data Augmentation and Normalization.}
    \begin{tabularx}{\columnwidth}{|X|c|c|c|c|}
        \hline
        \textbf{Experimental Setup} & \multicolumn{2}{c|}{\textbf{TTSA (\%)}} & \multicolumn{2}{c|}{\textbf{TOTA (\%)}} \\
        \cline{2-5}
        & \textbf{2-shot} & \textbf{32-shot} & \textbf{2-shot} & \textbf{32-shot} \\
        \hline
        w/o Data Norm. or Augm. & 61.2 & 73.4 & 38.67 & 42.8 \\
        \hline
        Data Normalization only & 77.3 & 83.3 & 45.8 & 56.8 \\
        \hline
        Data Norm. $\&$ Augm. & 79.69 & 83.68 & 49.9 & 60.4 \\
        \hline
        Prototypes Norm. + Data Norm $\&$ Augm. & 81.9 & 84.5 & 55.26 & 64.2 \\
        \hline
    \end{tabularx}
    \label{tab:protonet_conditions}
\end{table}

\subsection{Impact of Data Augmentation and Normalization }

We initiated a series of tests to determine the impact of normalization, data augmentation, and prototype normalization on our model's accuracy. Starting from a baseline accuracy of 38.67\% against datasets from different antenna configurations, each technique was applied sequentially. As detailed in Table \ref{tab:protonet_conditions}, these enhancements led to variations in accuracy, achieving accuracies of 55.26\% and 64.2\% for TOTA scenario with 2-shots and 32-shots respectively. The analysis focused on selected frames over a period of 500 samples of the total data ensuring that our subset was representative of the broader dataset, maintaining the integrity of our findings while reducing the complexity and duration of the experimental runs. Data normalization and prototype normalization proved most effective in improving accuracy, likely due to Prototypical Networks' classification mechanism, which operates similar to a k-nearest neighbors (KNN) approach, where maintaining scale and distribution consistency across prototypes is important.

\subsection{Analysis of K-Shots and Comparative Performance}

\begin{figure}[t!]
  \centering
  {\includegraphics[trim={0.85cm 0 2cm 0cm },width=0.41\textwidth]{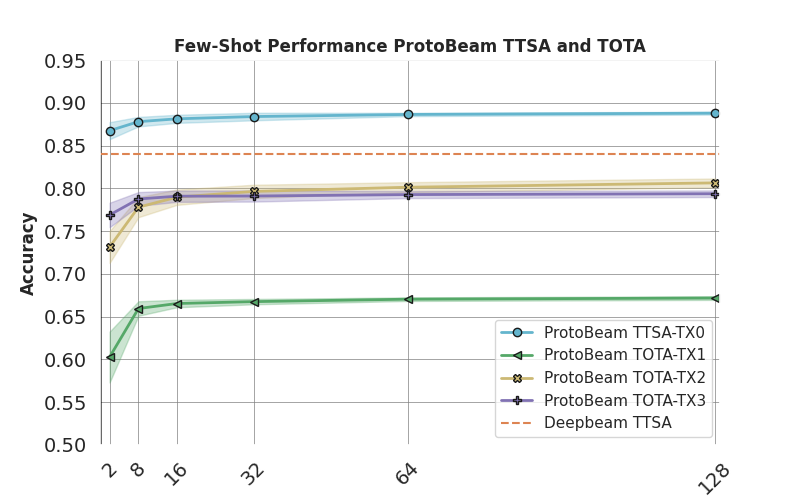} } 
  \caption{Accuracy vs. number of k (k-shot), comparing results from TTSA and TOTA to highlight model adaptability.}
  \label{kshot}
\end{figure}

Figure \ref{kshot} illustrates how the number of \textbf{k} in k-shot influences classification accuracy for our Prototypical Network. It presents separate performance curves for TTSA on antenna TX0, alongside performance when trained on TX0 and tested on different antennas: TX1, TX2, and TX3 (TOTA). In the TTSA scenario, accuracies of 87.36\% at 8-shot and 88.02\% at 64-shot are achieved. For TOTA scenarios, TX2 and TX3 exhibit accuracies around 79\%, and TX1 shows 68.28\%, demonstrating the model’s adaptability across different antenna setups. These results indicate that accuracy improves with an increase in \textbf{k} in k-shot up to a certain point, after which it plateaus, suggesting that additional samples beyond this threshold do not lead to substantial performance gains.

\textbf{Classification with a Tolerance:} 

\begin{figure}
  \centering
  {\includegraphics[trim={0.85cm 0 2cm 0cm },width=0.41\textwidth]{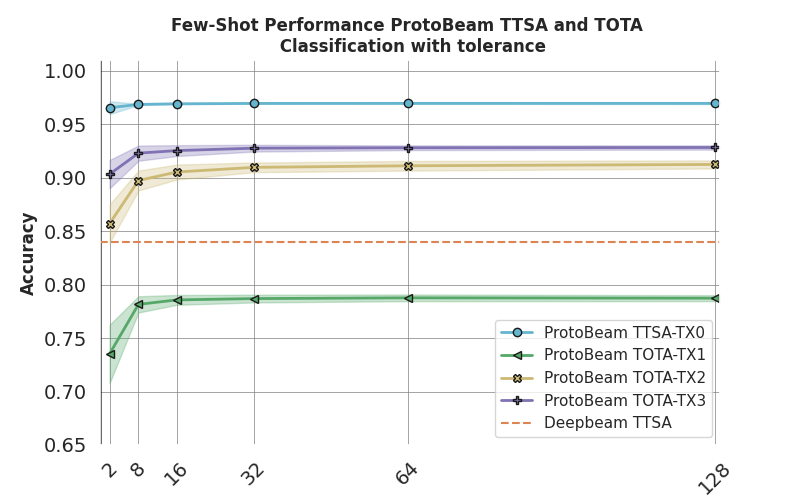} } 

  \caption{Accuracy vs. number of k (k-shot) with tolerance for both TTSA and TOTA experiments.}
  \label{stackedBar}
\end{figure}

To address observed misclassifications, we implemented a classification with tolerance strategy where the classification of the beam to be an adjacent beam is also treated as correct. This approach serves two main purposes: firstly, it allows for a margin of error to account for the model's inherent difficulty in distinguishing between neighboring beams, which tend to have similar characteristics. Secondly, it provides flexibility in considering misclassifications of neighboring beams as correct, recognizing that such errors may have minimal practical consequences in certain applications. The results with this tolerance are depicted in Figure \ref{stackedBar}. When applied at 16-shot, this method led to noticeable accuracy improvements: accuracy increased from 88.14\% to 96.91\% in the TTSA scenario and from an average of 74.11\% to 88.34\% in the TOTA scenario.

\subsection{Analysis of PCA Projections and Confusion Matrices} 

We analyzed the  PCA projections and confusion matrices for ProtoBeam to intrepret where misclassification happened. Figure 6(a) shows TTSA misclassifications mainly between neighboring beams like classes (5 \& 6), (11 \& 12), and (17 \& 18), where PCA illustrates close clustering. In contrast, Figure 6(b) for TOTA displays similar affected classes but tighter class prototype overlaps, highlighting the greater impact from domain shifts in these cases.

\begin{figure}[!tbh]
  \centering
  \subfigure[$TTSA$]{\includegraphics[width=0.47\textwidth]{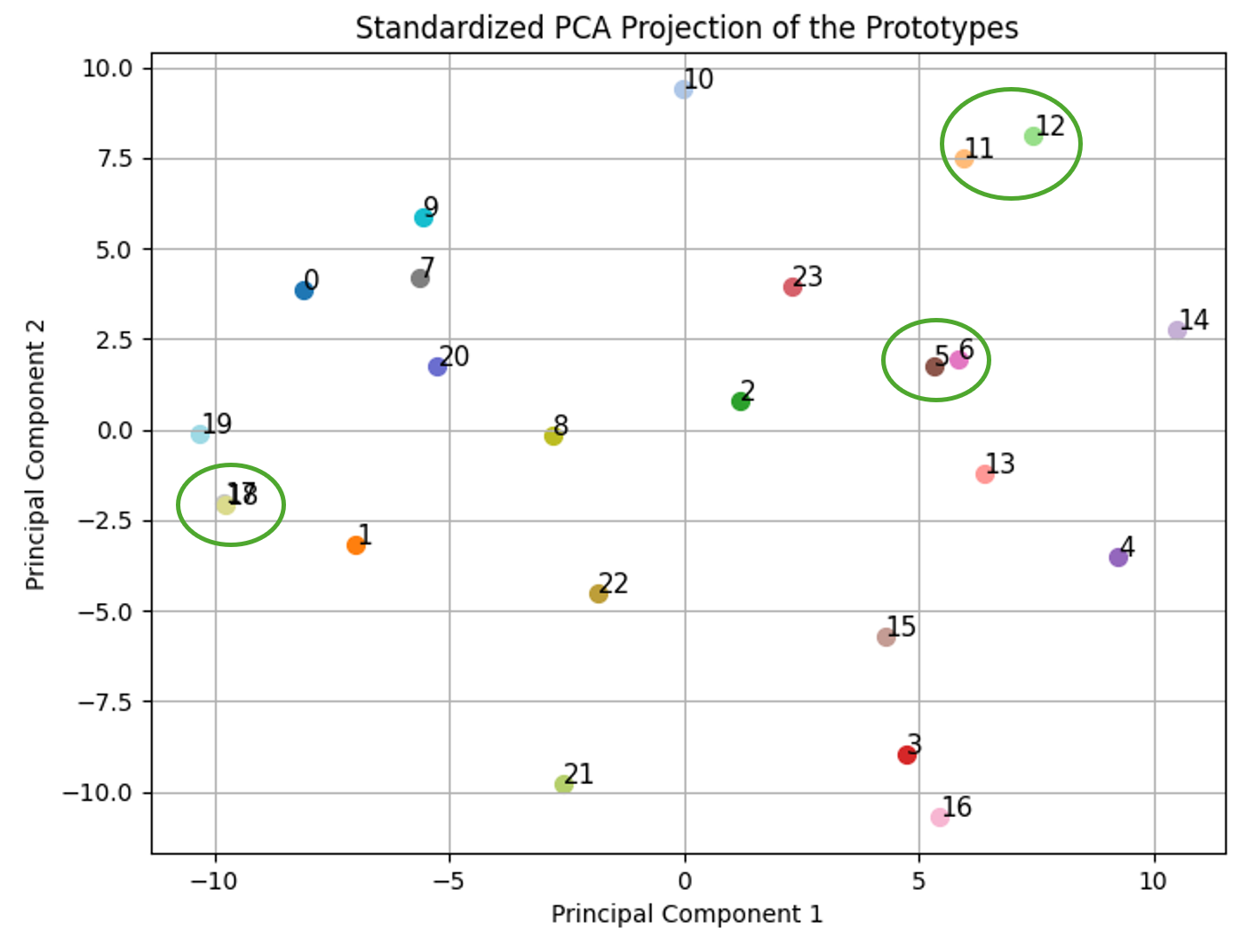}} \label{TTSA}
   \subfigure[$TOTA$]{\includegraphics[width=0.47\textwidth]{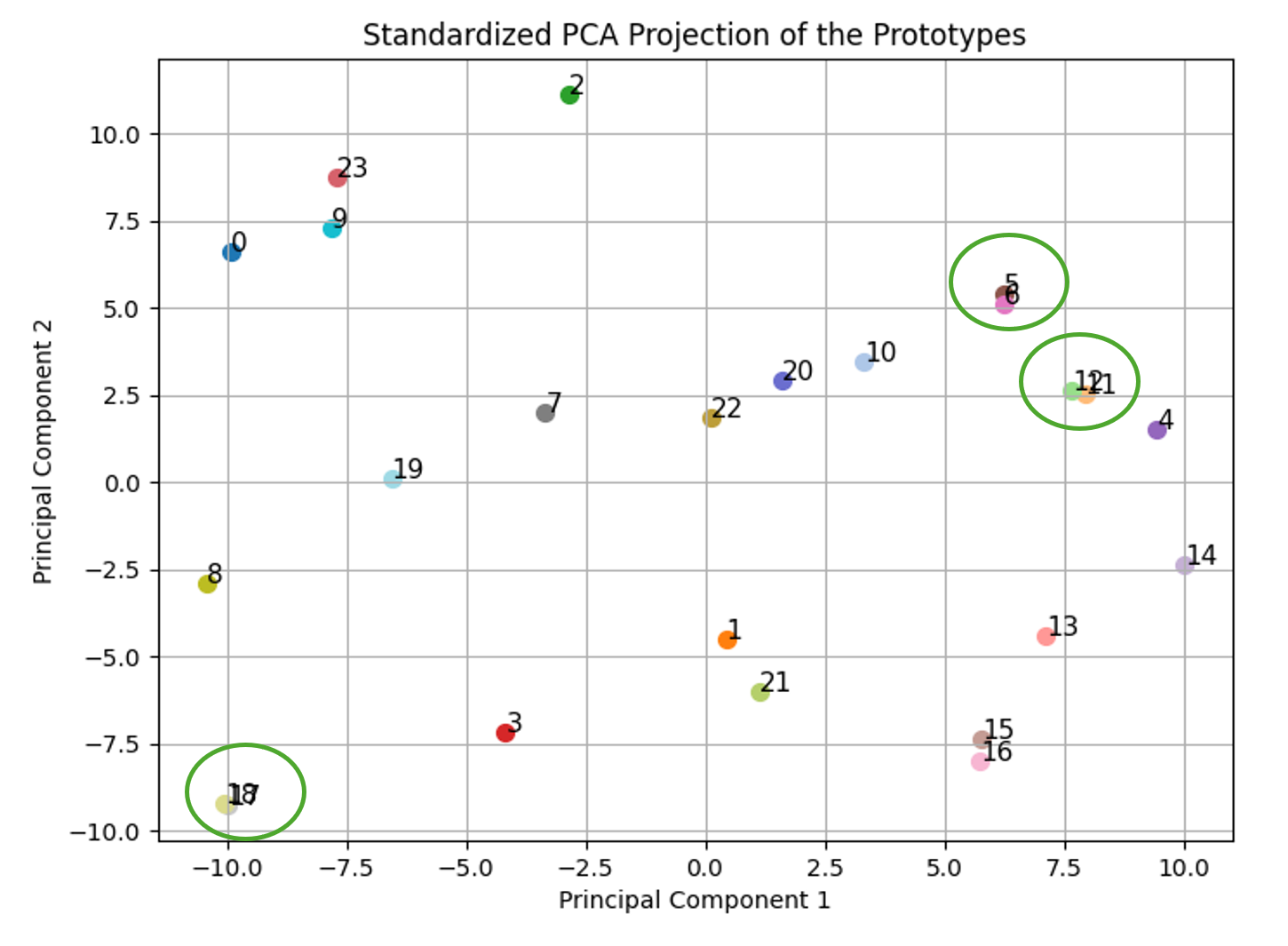}} \label{TOTA}

  \caption{Prototypical Network Analysis: PCA projections illustrate prototype proximity, highlighting overlapping classes.}
  \label{pca}
\end{figure}


\section{Conclusion}

The implementation of Prototypical Networks in mmWave beam classification has demonstrated substantial improvements, elevating classification accuracy from an average of 18.6\% to 74.11\% without the necessity for model retraining. This effectively addresses the complex challenges associated with domain shifts in wireless communication systems. This significant improvement demonstrates the potential of Prototypical Networks to enhance mmWave beam classification tasks with minimal reliance on additional samples from diverse antenna datasets. Future research will focus on refining our preprocessing techniques to enhance system efficiency and robustness by exploring zero-shot learning, particularly zero-shot Prototypical Networks. This approach aims to eliminate the need for samples from unseen or different antenna datasets by utilizing metadata.


\bibliographystyle{IEEEtran}

\bibliography{IEEEabrv,main}

\end{document}